\documentclass[twocolumn,showpacs]{revtex4}

\usepackage{graphicx}
\begin{document}
\title{Suppression of the $\gamma -\alpha$ structural phase
transition in Ce$_{0.8}$La$_{0.1}$Th$_{0.1}$
by large magnetic fields}

\author{F.~Drymiotis$^1$, J. Singleton$^1$, 
N.~Harrison$^1$, L.~Balicas$^2$, A.~Bangura$^3$,
C.H.~Mielke$^1$, Z.~Fisk$^4$, A.~Migliori$^1$,
J.L.~Smith$^1$ and J.C.~Lashley$^1$}
\affiliation{$^1$Los Alamos National Laboratory, 
Los Alamos, NM87545, USA\\
$^2$National High Magnetic Field Laboratory,
1800~E.~Paul Dirac Drive, Tallahassee, FL32310, USA\\
$^3$Condensed Matter Physics, Department of Physics, 
University of Oxford, The Clarendon Laboratory,
Parks Road, Oxford~OX1~3PU, United Kingdom\\
$^4$Department of Physics, University of California,
Davis, CA95616, USA
}
\begin{abstract}
The $\gamma-\alpha$
transition in Ce$_{0.8}$La$_{0.1}$Th$_{0.1}$
is measured as a 
function of applied magnetic field
using both resistivity and magnetization.
The $\gamma - \alpha$
transition temperature decreases
with increasing magnetic field, reaching
zero temperature at around 56~T.
The magnetic-field dependence
of the transition temperature
may be fitted using a model 
that invokes the field and temperature
dependence of 
the entropy of the $4f$-electron moments
of the $\gamma$ phase, suggesting that
the volume collapse in Ce and its alloys is
primarily driven by entropic considerations.
\end{abstract}


\maketitle
The element Ce has attracted considerable
theoretical and experimental interest over the
past fifty 
years (see {\it e.g.}
Refs.~\cite{lawson,koskenmaki,grioni,nikolaev,laegsgaard,held}
and references therein). 
One of the most fascinating
aspects of its behavior is the 
14.8~\% collapse in volume that occurs when the
face-centred-cubic (fcc) $\gamma$~phase
transforms into the fcc $\alpha$~phase on
cooling through $T_{\gamma \alpha} \approx 100$~K at ambient 
pressure~\cite{koskenmaki,laegsgaard,held}.
Although it is generally accepted that this
isostructural volume collapse is caused by the Ce~$4f$
electrons, a variety of models 
invoking different physical mechanisms have
been proposed. For example,
the {\it Mott transition model} proposes that
the $4f$ electrons behave as simple
band electrons in the $\alpha$~phase
but are localized 
on the Ce ions in the 
$\gamma$~phase~\cite{held,johansson1974,svane,johansson1995}.
This approach~\cite{johansson1974}
accounts qualitatively
for the reduction in magnetic susceptibility $\chi$
that accompanies the volume collapse~\cite{koskenmaki,koskimaki}.
However, such a picture
is certainly an over-simplification.
Neutron scattering~\cite{neutron}
and other experiments (see Refs.~\cite{liu1,liu2,dzero}
and citations therein)
suggest a large $4f$-electron spectral
weight on the Ce site in the $\alpha$ phase. 
This observation has been interpreted as evidence 
for ``localized'' $f$
electrons~\cite{laegsgaard,neutron}; however, a Ce
occupancy of $0.8 \pm 0.1$ could also be consistent with 
itinerant $f$ electrons within a narrow-band,
tight-binding picture.

The alternative {\it Kondo volume 
collapse model} proposes
that both the $\alpha$ and $\gamma$ phases possess
highly-correlated $4f$ electrons,
but with very different effective Kondo
temperatures~\cite{laegsgaard,neutron,liu1,liu2}.
In this scheme, the effective Kondo temperature
of the $\gamma$ phase is small,
({\it i.e.} less than $T_{\gamma \alpha}$);
in the $\gamma$ phase the properties of the 
$4f$ electrons will therefore
be almost indistinguishable from those of
localized ionic moments.
It is thought that the $\alpha$ phase has
a relatively large effective Kondo 
temperature by comparison~\cite{neutron,dzero},
causing the $4f$ electrons to be 
in the mixed-valence regime with significant
$spd$ and $f$ hybridization at low temperatures.
Consequently one might expect
the charge degrees of freedom
of $\alpha$~Ce to be describable in terms of
itinerant quasiparticles with a
large effective mass, a view that is
supported by recent optical data~\cite{Eb}.
Itinerant quasiparticles are 
preferable from an energetic standpoint
at low temperatures,
but the quasi-localized
$4f$ electrons of the $\gamma$ phase will be
favored on entropic grounds at elevated 
temperatures~\cite{laegsgaard,dzero}.

A magnetic field forms a useful probe of
the phenomena discussed above. Based on
the absolute size of $\chi$ in $\gamma$~Ce and
the large fall in $\chi$ at the 
volume collapse~\cite{neutron,koskimaki}, an applied field
will chiefly affect the energy and entropy of the
quasi-localized $4f$ electrons present
in $\gamma$~Ce; it will have a much smaller effect
on the $spd$ conduction electrons in $\gamma$~Ce
and on the itinerant quasiparticles  
of $\alpha$~Ce~\cite{dzero,Eb}.
In this paper, we have therefore applied 
steady magnetic fields of up to 27~T 
and pulsed fields of up to 57~T to samples
of polycrystalline Ce$_{0.8}$La$_{0.1}$Th$_{0.1}$.
We find that the field $B$ suppresses the
$\gamma - \alpha$ transition temperature, $T_{\gamma \alpha}$,
so that it extrapolates to zero at
a field $B_{\gamma \alpha}(T \rightarrow 0) \approx 56$~T.
This result, and the variation of $T_{\gamma \alpha}$
with $B$, are in agreement with a simple
model of the volume collapse based on 
entropy arguments~\cite{dzero}.

There are two reasons to choose
Ce$_{0.8}$La$_{0.1}$Th$_{0.1}$ rather
than pure Ce for such a study.
First, the Th content of the alloy
completely suppresses the dhcp
$\beta$ phase~\cite{grier,smith,manley}.
By contrast, in pure Ce at ambient pressure
the metastable $\beta$ phase (lifetime $\sim 10^4$~years)
is a considerable complication~\cite{koskenmaki}.
Successive thermal cyclings of pure Ce lead to
``contamination'' of the $\alpha$ phase
with varying amounts of $\beta$~Ce~\cite{koskenmaki,neutron}.
Moreover, $\beta$~Ce has a large
susceptibility, and the presence of even
tiny amounts in the $\alpha$ phase
sorely hinders 
magnetic measurements~\cite{koskenmaki,neutron}.
Second, the La content of the alloy
leads to a $T_{\gamma \alpha}(B=0)$
that is significantly lower
than that in pure Ce~\cite{grier,smith},
enabling the transition to be completely
suppressed by available magnetic 
fields.

The samples are prepared by arc melting the pure
metals in an argon atmosphere. The
Ce and La used are 99.99\% pure and
the Th purity is 99.9\%
(suppliers' figures). The raw materials
are melted together, and
then flipped and remelted
a number of times to
ensure sample homogeneity. 
The uniformity of the composition
is further ensured by annealing 
in vacuum for 8 days at $460^{\circ}$C.
Samples for the resistivity and magnetization 
measurements are spark
cut from the resulting ingots.
Electrical contacts are made using $50~\mu$m Pt
wires attached using spot welding.
Resistance data are recorded in
either a Quantum Design PPMS (data
from $0-14$~T) or a variable-temperature
probe within a 30~T Bitter magnet at NHMFL
Tallahassee (data from $0-27$~T); 
in both cases, the temperature sensor is
a calibrated Cernox resistor close to the
sample. The sample resistance is measured
using either a Linear Research LR700
bridge (Bitter magnet experiments) or the resistivity 
option of the PPMS; the two methods
are in good agreement.

\begin{figure}[htbp]
\centering
\includegraphics[width=8cm]{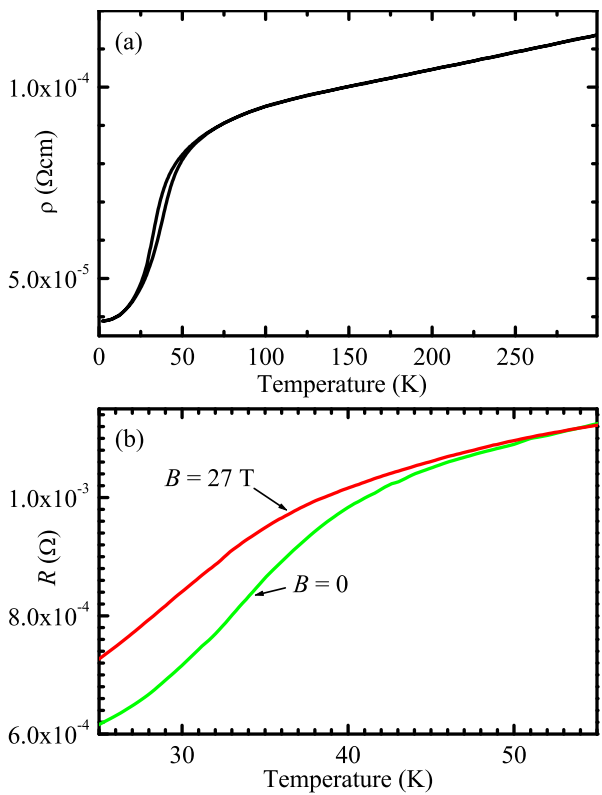}
\caption{(a)~Resistivity of Ce$_{0.8}$La$_{0.1}$Th$_{0.1}$
as a function of temperature at
zero applied magnetic field. The data
are recorded by sweeping the temperature 
at $\sim 1$~Kmin$^{-1}$ from
300~K to 2~K and then back to 300~K. Note
the hysteresis between down- and up-sweeps
of the temperature close to the transition.
(b)Resistance of a Ce$_{0.8}$La$_{0.1}$Th$_{0.1}$
sample versus temperature at
applied magnetic fields $B$ of 0 and 27~T.
In each case, the sample is been warmed above
200~K and then cooled at $\sim 1$~Kmin$^{-1}$
in the stated fixed field.}
\label{fig1}
\end{figure}

Fig.~\ref{fig1}(a) shows typical resistivity
versus temperature data with the sample in zero applied
magnetic field. The $\gamma - \alpha$
transition is visible as a
reduction in resistivity over a broad range
of temperature that starts at about
60~K; note that in agreement with previous
measurements of 
Ce$_{0.8}$La$_{0.1}$Th$_{0.1}$~\cite{smith},
there is hysteresis between data recorded
as the temperature falls and those taken as it rises.
Although the hysteresis is only clearly visible
close to the $\gamma - \alpha$ transition, there is
a large amount of irreversibility
and dissipation involved in the 
volume collapse; in common with other
studies of Ce alloys~\cite{grier,smith}, 
it is 
necessary to raise the sample temperature
to above 200~K 
before each sweep of the temperature 
down through the
transition to obtain consistent results. 
In what follows, 
we concentrate
on data acquired as the temperature {\it falls}
from 200~K through $T_{\gamma \alpha}$,
so that the transition is always from $\gamma$
to $\alpha$.

Various zero-field thermodynamic measurements of the 
Ce$_{0.8}$La$_{0.1}$Th$_{0.1}$ samples~\cite{lashley}
show that the $\gamma - \alpha$ transition occurs
at $T_{\gamma \alpha}(0) = 47 \pm 1$K. 
The corresponding point in
the resistance data is extracted using a variety
of methods ({\it e.g.} intersection of extrapolations
from above and below the transition, fitting
of ${\rm d}R/{\rm d}T$ to a Gaussian); in general,
these methods give values of $T_{\gamma \alpha} (0)$
within 1~K of each other
and within 1~K of the zero-field
thermodynamic measurements.

Fig.~\ref{fig1}(b) shows the effect of applied
magnetic field $B$ on the $\gamma - \alpha$ transition;
it is clear that the fall in resistance is displaced
to lower temperatures. Correspondingly,
the fitting
and extrapolation procedures mentioned above
give a transition
temperature of 
$T_{\gamma \alpha}(27~{\rm T}) = 40.9 \pm 0.7$~K.
Data acquired at several other fields
illustrate a similar trend;
{\it i.e.} the transition temperature is
suppressed by increasing applied field.

\begin{figure}[htbp]
\centering
\includegraphics[width=8cm]{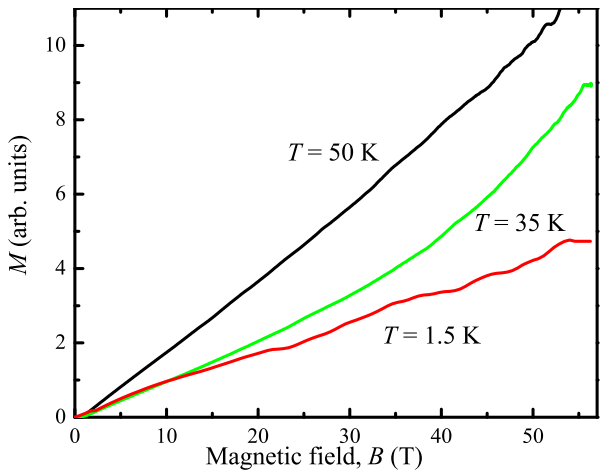}
\caption{Examples of pulsed-field 
magnetization data for a
Ce$_{0.8}$La$_{0.1}$Th$_{0.1}$ sample
at fixed temperatures $T$. 
Data corresponding to downsweeps
of the magnetic field are shown for $T = 1.5$, 35 and 50~K.
Before each shot of the pulsed magnet
the sample is warmed to above
200~K and then cooled at $\sim 1$~Kmin$^{-1}$
to the required temperature.}
\label{fig2}
\end{figure}

Magnetization measurements 
provide a complementary thermodynamic method for
examining the field dependence of the $\gamma - \alpha$
transition. Fig.~\ref{fig2} shows 
examples of such
data recorded at fixed temperatures
using an extraction magnetometer
inside a 57~T pulsed magnet at 
NHMFL Los Alamos~\cite{nhmfl};
before each measurement, the sample is
heated to 200~K and then cooled slowly 
($\sim 1$~Kmin$^{-1}$) to the
required temperature. 

Data for temperatures above $T_{\gamma \alpha}$
({\it e.g.} the $T= 50$~K data shown in
Fig.~\ref{fig2}) show a relatively
large susceptibility $\chi = {\rm d}M/{\rm d}H$, in agreement with
expectations for the $\gamma$ phase~\cite{neutron,koskimaki,grier}.
At very low temperatures, unambiguously
in the $\alpha$ phase ({\it e.g.} the
$T= 1.5$~K data in Fig.~\ref{fig2}),
the magnetization increases more slowly, consistent
with the small $\chi$  observed by 
others~\cite{koskimaki,grier}.
However, at intermediate temperatures (see the
$T= 35$~K data in Fig.~\ref{fig2}), there
is a distinct ``elbow'' or change in slope,
consistent with a transition
from a large $\chi$ (at high fields) to
a smaller $\chi$ (at low fields).
We associate this change in $\chi$
with a field-induced phase change from
$\gamma$ (high fields) to $\alpha$
(low fields); note that the data shown
are downsweeps of the field.
The field position of the transition
is taken to be the intersection of
linear extrapolations of
the low- and high-field gradients;
in this way, transition fields $B_{\gamma \alpha} (T)$
can be extracted from magnetization data at
a number of fixed temperatures $T$.

\begin{figure}[htbp]
\centering
\includegraphics[width=8cm]{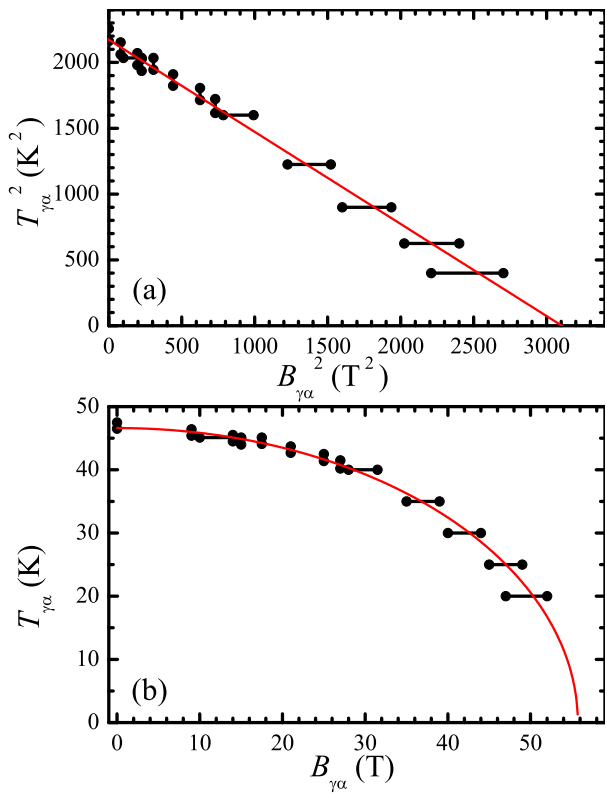}
\caption{Values
of $B_{\gamma \alpha}(T)$ from the
magnetization experiments at fixed
temperature $T$
(horizontal error bars)
and values of $T_{\gamma \alpha}(B)$
from resistivity experiments
at fixed field $B$ (vertical error bars)
plotted as $T^2$ versus $B^2$ (a)
and in linear units (b).
The line in (a) and the curve in (b)
represent Eq.~\ref{phase}
with $B_{\gamma \alpha}(T \rightarrow 0) = 56$~T
and $T_{\gamma \alpha}(B=0) = 46.6$~K.}
\label{fig3}
\end{figure}

The transition temperatures $T_{\gamma \alpha}(B)$
from the fixed-field
resistivity measurements and the transition
fields $B_{\gamma \alpha}(T)$ obtained from the 
fixed-temperature magnetization
experiments are summarized in Figs.~\ref{fig3}(a) and (b).
Data from the two techniques
may be distinguished as follows;
owing to the width of the transition,
the $B_{\gamma \alpha}(T)$ points from the magnetization
experiments have quite
a large field uncertainty~\cite{comment}. These data
are therefore plotted as horizontal error bars.
On the other hand, the $T_{\gamma \alpha}(B)$ values
from resistivity data
possess a relatively large 
temperature uncertainty, so that they
are represented by vertical error bars.
In spite of the differences between the
measurement techniques,
there is reasonable agreement between the
two sets of data shown in Fig.~\ref{fig3}. 
Note also that the data
lie on a straight line when plotted in
the form $B_{\gamma \alpha}^2$ versus
$T_{\gamma \alpha}^2$ (Fig.~\ref{fig3}(a)).

In order to understand the variation of
the $\gamma$ to $\alpha$ transition
with field and temperature, we turn to the two-phase
magnetic entropy model of Dzero {\it et al.}~\cite{dzero}.
This is based on the premise that
the low-temperature phase possesses
an energy scale characterized by
an effective Kondo temperature $T_{{\rm K} \alpha}$
that is much larger than the corresponding
energy scale for the high-temperature phase,
$T_{{\rm K}\gamma}$, {\it i.e.}
$T_{{\rm K} \alpha} \gg T_{{\rm K} \gamma}$.
This view is supported by variety 
of experiments that suggest 
$k_{\rm B}T_{{\rm K}\alpha} \sim 80-170$~meV
and $k_{\rm B}T_{{\rm K}\gamma} \sim 5-8$~meV
(see Refs.~\cite{laegsgaard,liu1,liu2,neutron,dzero} 
and references therein).
The large value of $T_{{\rm K}\alpha}$ means
that the free energy $F_{\alpha}(B,T)$ 
of the $\alpha$ phase will vary
only slowly with $B$.
By contrast
the quasi-localized $4f$ electrons of
the $\gamma$ phase will couple strongly to the applied
field, producing a large negative contribution to the 
$\gamma$ phase's free energy, $F_{\gamma}(B,T)$.
This distinction is emphasized~\cite{dzero} by writing
$F_{\gamma} = E_0 -TS(B,T)$, where $E_0$
indicates the contribution of the intinerant
$spd$ electrons and $S(B,T)$ is the entropy
associated with the $4f$ multiplet of angular
momentum $J$, Land\'{e} $g$-factor $g_J$;
\begin{equation}
TS(B,T) = -T \log_{\rm e}\left(\sum^{J}_{m_J =-J}
\exp[-\frac{g_J \mu_{\rm B}Bm_J}{T}]\right).
\label{entropy}
\end{equation}
The phase boundary is defined by the equivalence
of the free energies, $F_{\gamma}(B,T) = F_{\alpha}(B,T)$.
Owing to the fact that both $F_{\alpha}(B,T)$ and
$E_0$ will vary relatively slowly with $B$ and $T$
compared to $TS(B,T)$, the approximate condition
for the phase boundary becomes $TS(B,T) \approx$~constant.
With this constraint,
straightforward manipulation of 
Eq.~\ref{entropy} (see Ref.~\cite{dzero})
yields the following equation
relating the fields and temperatures on the 
$\gamma - \alpha$ phase boundary;
\begin{equation}
\left(\frac{B_{\gamma \alpha}}{B_{\gamma \alpha}(T\rightarrow 0)}\right)^2
+\left(\frac{T_{\gamma \alpha}}{T_{\gamma \alpha}(B=0)}\right)^2
\approx 1,
\label{phase}
\end{equation}
{\it i.e.} a plot of $B_{\gamma \alpha}^2$ versus 
$T_{\gamma \alpha}^2$ should yield a straight line.

The straight line in Fig.~\ref{fig3}(a) is a fit
of the data to Eq.~\ref{phase} with
$B_{\gamma \alpha}(T \rightarrow 0)$ and $T_{\gamma \alpha}(B=0)$
as adjustable parameters.
The values obtained were 
$B_{\gamma \alpha}(T \rightarrow 0) = 56\pm 1$~T~\cite{comment}
and $T_{\gamma \alpha}(B=0) = 46.6 \pm 0.5$~K.
A consistency check of these values can be
obtained by again setting $TS(B,T)$ equal
to a constant and examining Eq.~\ref{entropy}
in the limits $B=0$ and $T\rightarrow 0$;
this produces
\begin{equation}
\frac{k_{\rm B}T_{\gamma \alpha}(B=0)}{\mu_{\rm B}B_{\gamma\alpha}(T\rightarrow 0)}
= \frac{g_J J}{\log_{\rm e}(2J+1)} \approx 1.20,
\label{ratio}
\end{equation}
where we have inserted the known values~\cite{koskimaki,neutron}
$J=\frac{5}{2}$ and $g_J=\frac{6}{7}$ for the quasi-localized
$f$ electrons of $\gamma$~Ce.
Using the fit parameters derived from
Fig.~\ref{fig3}, the experimental ratio 
$k_{\rm B}T_{\gamma \alpha}(B=0)/\mu_{\rm B}B_{\gamma \alpha}(T \rightarrow 0)$
is 1.24, within a few percent of the prediction of Eq.~\ref{ratio}.

It is interesting to note that a rather simple
model~\cite{dzero} produces a 
internally self-consistent 
description of these data.
The model, in effect, neglects the
field and temperature dependences
of {\it both} the $spd$ electrons of the $\gamma$ phase,
and the itinerant, hybridized $spdf$
state~\cite{Eb} 
of the $\alpha$ phase, treating only the
free energy contributed by the
well-defined moments present in the $\gamma$ phase.
Hence, the success of the model
suggests that the
$\gamma - \alpha$
transition is driven mainly by the fact
that entropy considerations
favor $f$ electrons that are effectively localized
at high temperatures and fields
(and, one might add, at low pressures~\cite{laegsgaard}).
The delocalization of the $f$
electrons manifested as the 
itinerant quasiparticle behavior~\cite{Eb} 
of the $\alpha$ phase
is energetically favorable
from zero-point-energy considerations, but it
is costly on entropic grounds.
Thus, $\alpha$~Ce is stable at low 
temperatures and fields (and at high pressures).
Finally, we remark that it has been
argued on the basis of a qualitative
assessment of phonon spectra
that the entropy of the $4f$ moments
plays a dominant role in 
determining the structural phase of
Ce$_{0.9}$Th$_{0.1}$~\cite{manley}.
The success of the magnetic entropy model
in the current work adds a {\it quantitative}
justification to this argument.

In summary, we have measured the $\gamma-\alpha$
transition in Ce$_{0.8}$La$_{0.1}$Th$_{0.1}$
as a function of applied magnetic field
using both resistivity and magnetization.
The transition temperature is suppressed
by increasing magnetic field, extrapolating
to absolute zero at around 56~T.
The magnetic field-temperature 
phase boundary is adequately fitted by
a simple model~\cite{dzero} 
of the field and temperature
dependence of 
the entropy of the localized $f$-electron moments
in the $\gamma$ phase. This suggests that
the volume collapse in Ce and its alloys is
primarily driven by entropic considerations.
Many substances undergo structural
phase transitions which are thought to involve
$f$ electrons changing from ``localized''
to itinerant behavior ({\it e.g.} Plutonium~\cite{pu}).
On the basis of the work reported in
the present paper, it seems likely 
that the application of
high magnetic fields to such substances
will yield valuable information 
about the role of the $f$ electron
system in stabilizing
the various structural phases.

This work is supported by the U.S. Department
of Energy (DOE) under Grant No. LDRD-DR 20030084
``Quasiparticles and Phase Transitions in High 
Magnetic Fields: Critical Tests of Our Understanding
of Plutonium'' and by the National Science 
Foundation (NSF) (grant DMR-0433560).
It is also sponsored by the National Nuclear Security
Administration under the Stewardship Science Academic
Alliances program through DOE grant DE-FG03-03NA00066. 
Part of this work was carried
out at the National High Magnetic Field Laboratory,
which is supported by NSF,
the State of Florida and DOE. We thank Lev Gor'kov,
Alex Lacerda and Peter Littlewood
for enthusiastic and helpful comments and are grateful to
Stan Tozer for experimental assistance.

\end{document}